# Volume Fabrication of Quantum Cascade Lasers on 200 mm-CMOS pilot line

JG Coutard[1], M Brun[2], M Fournier[1], O Lartigue[1], F Fedeli[1], G Maisons[2], JM Fedeli[1], S Nicoletti[1], M Carras[2] and L Duraffourg[1*]

[1] Univ. Grenoble Alpes, CEA, LETI, F38054 Grenoble.
[2] mirSense - Centre d'intégration NanoINNOV, Bâtiment 863, 8 avenue de la Vauve F91120 Palaiseau

* Laurent.duraffourg@cea.fr

**Abstract**

The manufacturing cost of quantum cascade lasers is still a major bottleneck for the adoption of this technology for chemical sensing. The integration of Mid-Infrared sources on Si substrate based on CMOS technology paves the way for high-volume low-cost fabrication. Furthermore, the use of Si-based fabrication platform opens the way to the co-integration of QCL Mid-InfraRed sources with SiGe-based waveguides, allowing realization of optical sensors fully integrated on planar substrate. We report here the fabrication and the characterization of DFB-QCL sources using top metal grating approach working at 7.4 µm fully implemented on our 200 mm CMOS pilot line. These QCL featured threshold current density of 2.5 kA/cm² and a linewidth of 0.16 cm$^{-1}$ with a high fabrication yield. This approach paves the way toward a Mid-IR spectrometer at the silicon chip level.

**Introduction**

These two last decades, hybrid photonics circuits using both silicon based materials and III-V materials have been successfully developed for data communication and are a complementary technology to advanced CMOS for high performance computers too. In 2006, Soref suggested to consider a similar technological approach to make photonics circuits dedicated to the mid/long wave infrared region [1]. This IR-wavelength range enables to address multiple applications from free-space data communication [2], optical countermeasures [3], IR imaging of biological tissues [4] to spectroscopy [5]. So far, the detection of chemical species in solid [6], liquid [7] or gas mixtures [8] remains the most popular application that drives the technological development. In fact, the spectral range between 3 µm and 12 µm corresponds to the first harmonic resonance between rovibrational energy levels of the most chemical species, leading to absorption cross sections that are order of magnitude stronger than overtone in the near IR for instance. This is especially true for light molecules in gaseous phase. This unique feature enables to detect a great number of gasses with extreme sensitivity and selectivity. A limit of detection lower than 1 part per billion (ppb) and unequivocal identification of chemical species can typically be reached through MIR absorption spectroscopy techniques [9].

Even if chemical detection using diode lasers has been developed since the mid-1960s, it is with the advent of unipolar sources based on multiple quantum well stacks that spectroscopic sensing in the MIR wavelength band has become commercially available. With the recent progress in Quantum Cascade Laser (QCL) [10], [11] and Interband Cascade Laser (ICL) technology, compact and reliable MIR light sources are now readily available [12]. In particular, Distributed FeedBack (DFB) sources allow the selection of specific emission wavelengths to target the detection of chemicals of interest [13], [14]. With these sources, a novel generation of sensing tools suitable for real-time in-situ detection of trace element gasses is now available. With the advent of MIR Si photonics, a novel class of integrated components has been developed allowing the integration at chip level of the main building blocks required for chemical sensing, i.e. the source, a photonics integrated circuits (PICs) and the detector [15]. At wavelengths around 4.8 µm, Spott et al. developed a silicon on SiN waveguide coupled with a bonded QCL material. DFB lasers with threshold currents as low as 80 mA and threshold current densities below 1 kA/cm2 emitted more than 200 mW from a polished III-V/Si facet, and operated in pulsed mode up to 100 °C [16]. They rely on InP type fabrication using III/V-manufacturing lines on two inch format.

Such technological approaches cannot satisfy volume markets. Today manufacturing price may be estimated around 1k€ per laser after fabrication, test, sorting and bonding of the die on its holder. To date, even if III-V technologies are sufficient to meet the needs of niche markets they cannot tackle the QCL volume manufacturing challenges: implementation of automatic testing procedures at the wafer level, improvement of the reproducibility of electro-optical features, simplification of the sorting operations and implementation of a quality-control system. These elements enable bringing to market at a suitable manufacturing cost, most probably around few € with volume largely higher than 100 kunits per year.

So far, these challenges have not been properly addressed and prevent from a large development and dissemination of these lasers beyond scientific community and niche applications. In this paper, we present an original approach for the Mid-IR based upon the use of microelectronic tools to realize the fabrication of QCL on 200 mm Si wafers. Doing so, we are able to dramatically increase the laser reproducibility, while preserving the same performances as those reached on InP and to set-up automatic test procedure for reducing or even removing the sorting of lasers. This paves the way to the manufacturability of low cost devices suitable for numerous applications from single laser to complete analysis system of chemicals, biologicals at liquid or gas phases. This work address a fundamental building block for the co-integration of QCL array with suitable MIR PICs and PhotoAcoustic detectors for making a spectrometer fully miniaturized at chip level.

This paper reports on the design, fabrication, and characterization of DFB-QCL single sources and arrays made on 200 mm Si wafers and compares their performances with QCLs made on InP substrate, whose designs are similar to those used in this work. For the sack of clarity, here following we will mainly focus on the devices emitting at 7.4 µm, even if similar experimental results have been obtained at 4.5 µm.

**Design**

QCLs are commonly based on III-V materials depending on the chosen wavelength range. Nowadays, most of the QCL emitting in the MIR region (in particular between 3 µm and 11 µm) are made from a stack of InGaAs / AlInAs layers on InP substrate. These heterostructures are significantly more efficient than the GaAs / AlGaAs stacks and have the optical power record by reaching up to few watts [17], [18] in continuous mode and at room temperature. InAs based QCL are also good candidates in particular for short wavelengths around 3 µm. More recently, Baranov and co-workers have demonstrated InAs / AlSb QCL emitting at 15 µm wavelength with a low threshold current density at room temperature [19].

The QCL developed within this work will be used for the chemical analysis in the 4 - 10 µm wavelength range. The design has been derived from previous stacks developed a couple of years ago and specifically designed to operate in the 7.4 µm emission wavelength bands. The laser heterostructure is a standard lattice-matched GaInAs and AlInAs multilayers grown on InP. Our active region is composed of 25 elementary periods, which are composed of a 3-quantum-wells and a superlattice resonant injection regions. For more details on the heterostructure, the reader can refer to [13].

The electroluminescence spectrum of the epitaxy layers, measured at 80 K, showed that the emission band is actually centered at 1400 cm$^{-1}$ (7.1 µm) with a full width at half maximum, $FWHM = 14.4\ meV$.

The multi-layer stack designs were however modified from DFB-QCL sources originally done on InP to take into account the constraints set by CMOS processing tools. The bottom contacts of the lasers are in particular no longer on the backside of the chips, but reported aside from the ridge of the laser introducing doped layers below the active region.

Three lengths of ridge have been investigated ($L_{ridge} = 1, 2, \text{and } 4\ mm$). We designed four ridge widths per length ($w_{ridge} = 6, 8, 10 \text{ and } 12\ \mu m$) for studying the current density threshold and the related overlap with the active region that should degrade the output power at same injected current density level. The nominal ridge geometry is 10 µm-wide, 2 mm-long. It corresponds to a good compromise between gain and losses and enables to have a single spatial mode. This width is adjusted to increase the losses of the higher order modes to the benefit of the first spatial mode. The DFB-QCL requires a wavelength selection mechanism, which is done with a metal surface grating according to the approach coupling a surface plasmon-polariton mode and the propagative modes [13].

**Technology and fabrication**

To fabricate the laser on the 200 mm CMOS pilot line, we directly bonded a 2 inch InP wafer with the active layers on a 200 mm Si wafer via a thin oxide layer. This process step is a now mature technology but it requires high-quality defect-free epilayers. This key process step has been addressed first by the study of the impact of the surface quality of the InP wafers with epitaxial QC stack from different providers. The bonding is very sensitive to defects and requires a low roughness and a low bow. If the availability of defect-free epitaxial layers are primordial, the quality of the InP substrates is also critical. Provided that the surface defects are low enough on the epitaxial wafers, we achieved direct bonding of more than 90% of surface on oxidized 200 mm Si wafer. The InP substrate was first grinded and then fully removed by chemical etching of the different sacrificial layers in order to leave a perfect upper surface for further processing. It is worth to mention that a single Si wafer can receive up to four 2 inch wafers so that the number of devices fabricate on a single wafer can be increased from 2700 up to 10000 single lasers per wafer.

This QCL manufacturing is fully compatible with standard microelectronics processes as it directly exploits the top metal grating technology, which does not require any further process on the InP wafer before the molecular bonding, ensuring the high volume rate. The implementation of QCL fabrication process on silicon benefit from the specific DFB technology developed a few year ago [20], [16] and routinely implemented on InP [13], [21].

The manufacturing of DFB QCL is realized through a grating in the III/V stack exhibiting a periodic index variation. This approach implies regrowth of III-V material after the grating etching [10]. Today, this regrowth is not possible in a CMOS-like manufacturing of QCL on silicon wafer. To overcome this issue, the wavelength selection, presented in [16], is achieved on a silicon/nitride waveguide. This approach requires to couple the QCL cavity to a silicon/oxide waveguide, which exhibits two main drawbacks. It complicates the design to have an efficient coupling inside the laser cavity, and reduces the useful wavelength range due to the presence of oxides in the waveguide. Oxide usually exhibits high losses above 4 µm mid IR wavelength (likewise SiN above 7 µm). The top metal grating approach in double trench does not require regrowth and has therefore provided the basis for the manufacturing process that we present in this paper.

A typical silicon wafer with the active 2" area including the QCLs is presented in Fig. 1 a. The main challenge encountered in the fabrication of QCL within a 200 mm CMOS compatible facility was the development of suitable processes matching the etching of thick layers up to 10 µm and high resolutions down to 300 nm. For deep UV lithography at 248 nm, the thicknesses of the resists, the coverage of steep and steps depth are the key points compared to the fabrication of Near-IR lasers heterogeneously integrated on Si. The partial etch of InP and InGaAs layers combined with precise stopping level are the challenges for the etching steps. The deposited material should be optically and chemically compatible with QCL structure. We used SiN for fabricating the hard mask for InP etching as well as for the deposition of conformal thick layers on steep edges. Thick conformal gold layers are used for the DFB operation and for getting good contacts. Finally, the complete process has been performed in our MEMS 200 mm Si platform enabling the delivery of numerous DFB QCL lasers and QCL arrays. In this work, we manufactured five wafers, three centered on 7.4 µm emission wavelength and two centered on 4.5 µm emission wavelength. A 200 mm Si wafer with components (Fig. 1 a), a lithographic 22 x 22 mm² field and a SEM zoom-in of a single QCL are shown in Fig. 1 b&c. The performance of lasers depends not only on the quantum well stack but also on the morphology of the laser ridge and the DFB filter. SEM characterizations were performed to verify the geometry of the ribbon and the quality of its edges. The expected widths were larger by 1.4 µm. The manufacturing accuracy is in the order of +/-100nm over the entire center of the 200mm wafer (Fig. 1a)). The roughness of the etching flanks is less than 100nm (RMS). The ribbon edges are metallized with gold to increase the losses of higher order optical modes and to have a single mode laser. It should be noted that the metallization of the back mirror allows a 95% reflectivity and that an anti-reflective layer (SiN) ensures a 95% transmission on the output facet.

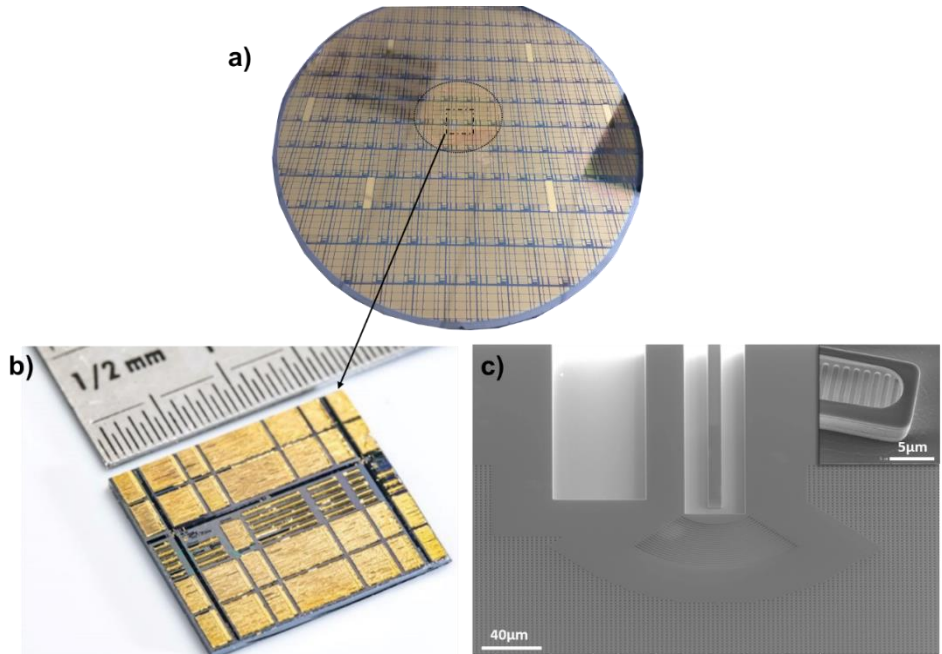

Figure 1: a) 200 mm Si wafer with QCL components (the dotted circle refers to plate 2" InP & the dotted square refers to a laser field) – b) Field of (22 mm x 22 mm) of single QCL devices and QCL arrays – c) SEM picture QCL laser – inset: Zoom in of the DFB grating (at the end of the ridge tip)

**Electro optic characterizations**

As aforementioned, the fabrication of QCLs on 200 mm substrate enables to use large-scale characterization tools. Thus, 2700 devices (corresponding to single QCLs and QCL arrays) were measured at wafer-level on an automatic 200/300 mm test prober to determine the emission threshold levels. To do so, $V(I)$ and $P(I)$ characteristics were systematically measured using synchronized IR detector (VIGO system, Poland) to measure the output light emission. The measurements were performed in pulsed mode at low duty cycle (3%) and at 17 °C to avoid any self-heating effect. The pressure exerted on the electrical contact pads as well as the positioning of the prober tips remain delicate and may modify the series resistance (typically in the order of 1 Ω or less). Calibration of our prober has shown an average contact resistance of a few tens of mΩ (and a dispersion in the order of 1 mΩ), which is negligible compared to the differential resistance of our QCLs that are typically around 40Ω. We extracted the repartition of threshold current densities, $J_{th}$, per wafer, which enables the selection of functional dies and the evaluation of the fabrication yield. In particular, we computed the average threshold currents and the standard deviations for each geometry of ridges.

In Figure 2, we have represented the threshold current density as a function of the length and width of lasers. Each point corresponds to the average value of the current density and its standard deviation computed over 225 identical lasers. Figure 2a) corresponds to measurements made on one typical wafer. Figure 2b) shows the averages and standard deviations computed from the cumulative measurements done on two wafers according to the ridge length. For reference, we have superimposed the typical (non-statistical) threshold current densities measured on lasers manufactured on InP. For the shortest lasers (1mm), the threshold current density is around 4.6kA/cm² and drops for the longest lasers (4mm) down to 2.5 kA/cm². The threshold current decreases rapidly with the length of the DFB grating, whose reflectivity tends towards 1 from a value close to or greater than 4mm. The losses of the laser ribbon becomes predominant from this length. The threshold current densities exhibited by QCL fabricated on InP substrate remain lower (2.6 kA/cm² & 1.9kA/cm² for 2mm and 4mm respectively). As reported in the fabrication section, the bottom contacts reported aside from the ridge of the laser requires doped layers below the active region, which may induce additional losses, which tend to increase the threshold current.

Figure 2c) shows the average values and standard deviations as a function of width (values from SEM observations). We reported two sets of measurements corresponding on two different wafers. The orange circles are for the typical wafer (same as in Fig 2a)) since the blue ones are for a second wafer from the same fabrication batch. There is no clear trend in the variation of the threshold current with the width. For the lengths of 1mm and 2mm, it nevertheless seems that lasers around 10µm wide have the lowest current densities. This observation appears to be consistent with the initial sizing of the laser, whose theoretical optimal width was set at 10µm for the stack considered. The values from the two wafers remain very close, in particular for 1mm and 2mm long lasers. The 4mm long lasers of the second wafer show a larger dispersion in the measured current densities.

The relative standard deviation $\sigma_{J_{th}}/\overline{J_{th}}$ is about 3% except for the 4mm long lasers of the second wafer. This weak dispersion demonstrates the pretty good robustness of our technology. To go further, it is interesting to compare the disparity of the threshold current values with the dimensional dispersion related to our lithography/entching processes, used to structure the laser ribbon and DFB grating. The tolerance of the manufacturing process is better than 100nm over a ridge width and does not induce the relative dispersion of few percent observed in Fig. 2. Moreover, the DFB design uses a coupling of the surface plasmon mode at the metal / dielectric interface with the guided mode [13], [20]. This approach enables a coupling strength of the grating and losses that are almost constant over a wide range of etching depths [20]. We estimated that the acceptable tolerance depth is +/-100nm. The variation of our etching process remains below this limit (typically +/-50nm measured on few samples) what should not significantly influence the threshold current.

With this systematic electro-optical characterization, we estimated a yield of 98% of functional lasers per wafer (with similar electro-optical features).

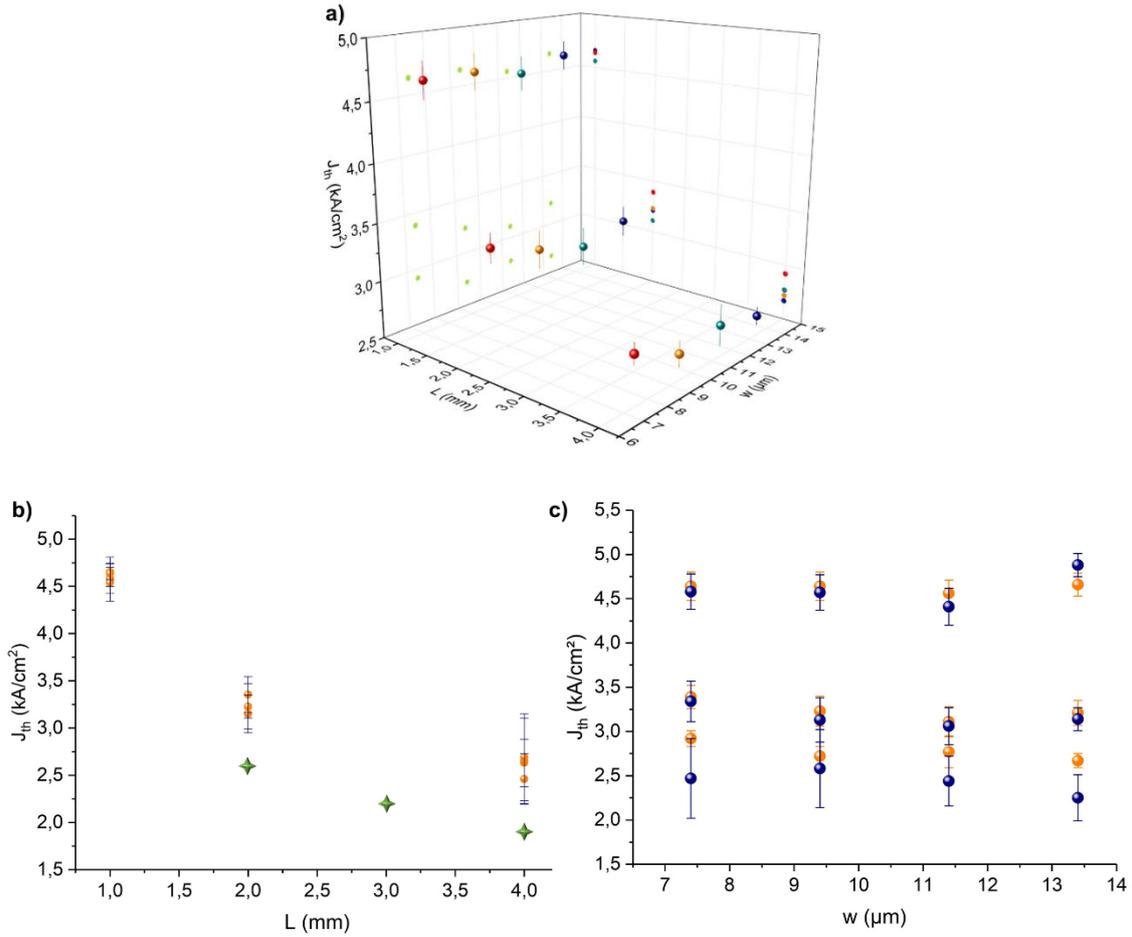

Figure 2: Current density threshold: a) average and standard deviation values with the length and the width (225 dies per geometry) for one typical 200 mm-wafer at 7.4 µm emission – b) average and standard deviation values with width (orange circles: wafer 1, blue circles: wafer 2) – c) average and standard deviation values as a function of length for the two silicon wafers (orange circles) and for lasers made on InP (green stars)

After these first characterizations, the wafers were diced into discrete components (2700 dies / wafer). $P(I)$ and $V(I)$ were once again measured according to the applied current through the laser to extract $J_{th}$ and the maximum current density $J_{max}$. These measurements were performed at four different operating temperatures (from 15°C to 45°C) at 1.5%-duty cycle. When the current density reaches $J_{max}$, the injection band level becomes higher than the transition level and the Stark rollover effect appears. In this regime, the optical power drops with the current. From the curve at different temperature, we estimated the variation of $J_{th}(T)$ and extracted the characteristic temperature $T_0$ according to (1).

$$J_{th}(T) = J_0 e^{\frac{T}{T_0}} \qquad (1)$$

In the figure 3, we present typical $P(I)$ and $V(I)$ curves obtained for the nominal geometry (theoretical ridge width= 8µm, measured width = 10.4 µm, ridge length = 2 mm). As shown in inset of Fig. 3, $T_0$ is close to 176 K, which corresponds to quite common value obtained on InP substrate. This demonstrates that both the bonding process of quantum well stacks on Si wafers and the ridge /DFB structuration using CMOS compatible tools do not degrade the electro-optical performances of the QCL. The roll-off currents are quite common too. These characterizations have however been done at low duty cycle preventing from large self-heating occurrences and no systematic measurements of the heat diffusion through the Si substrate has been made so far. Further measurements at the die level are in progress to better characterize the QCL when working in pulsed mode up to 15%-duty cycle. In this case, the average optical powers would be between 5 mW and 10 mW.

A differential optical power $dP/dI$ of ~83 mW/A can be easily estimated from measurements presented in Fig. 3 considering the duty cycle of 1.5%. For comparison, typical differential powers that can be found with commercial lasers on InP are around few 100mW/A in similar conditions (in pulse mode and at 15°C) [22].

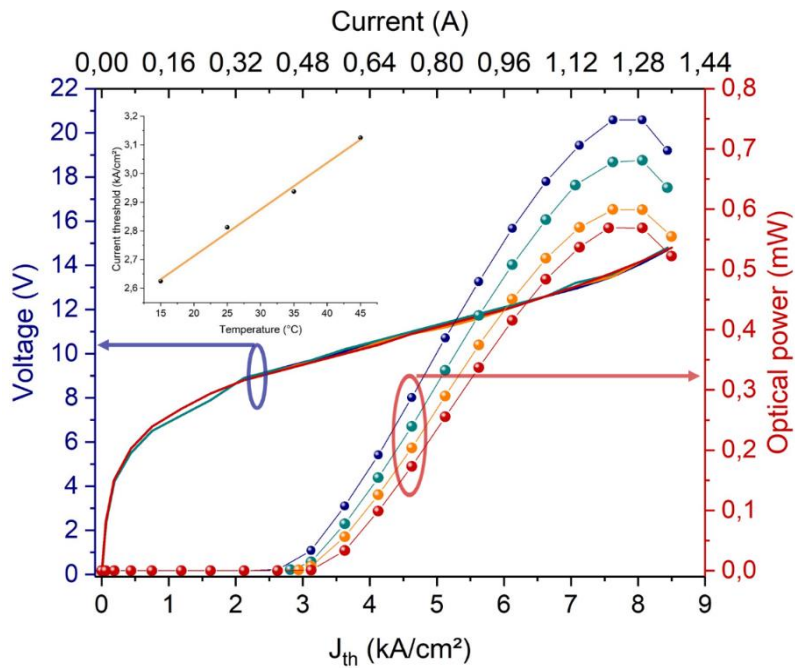

Figure 3: Typical characteristics $P(I), V(I)$ of QCL lasers at 7.4µm emission wavelength (ridge width=8 µm, ridge length=2mm) at four temperatures: blue: 15 °C, green: 25 °C, orange: 35 °C and red: 45 °C – inset: $J_{th}(T)$ vs. temperature

We also report the normalized typical emission spectra of an array of QCLs working in the 7.4 µm wavelength range for the nominal geometry in Fig. 4a). The output power was measured operating the lasers at the same condition (applied voltage = 9.9 V, pulsed mode with a 50 ns-pulse duration and 100 kHz-repetition rate (*i.e* 0.5%-duty cycle), operating temperature T=21 °C. The spectral resolution of our Fourier-transform infrared spectrometer (Thermofisher Scientific Nicolet IS-50) is 0.0125 cm$^{-1}$ (4375 points over the spectral range). Apart from sources 2 and 6, the devices showed single-mode emission and 25 dB side mode suppression ratios over the working wavelength range (see Fig 4b) insert). The emission wavenumbers extracted from this measurement are in good agreement with the expected wavelengths defined by the DFB design, as shown in Fig 4 b). In the figure 4c), we plot the intensity emitted by a QCL at the middle of the array (*i.e.* QCL12) with the wavenumber. The maximum power density is 199 µW at 1357.6 cm$^{-1}$. The optical power integrated over the entire emission spectrum corresponds to 18 µW (@0.5% duty cycle). This value is coherent with the power measured in Fig. 3 at 10 V at 1.5%-duty cycle (i.e. 54 µW nearby the threshold). In this operating condition, the typical *FWHM* is 0.16 cm$^{-1}$ with a current pulse duration of 50ns and a repetition rate at 100 kHz. The *FWHM* is in particular affected by self-heating phenomenon that broadens the linewidth (*FWHM* of a free running QCL is between 1MHz-10MHz in CW mode with an efficient thermal drain) [23]. Other technical noises of the set-up results also in a spectral broadening, and the self-heating contribution cannot be properly quantified with our current set-up. The emission spectrum even widens rapidly for pulses exceeding 100ns up to 1 cm$^{-1}$ (*FWHM*) for 100ns-pulse duration and 13V of applied voltage.

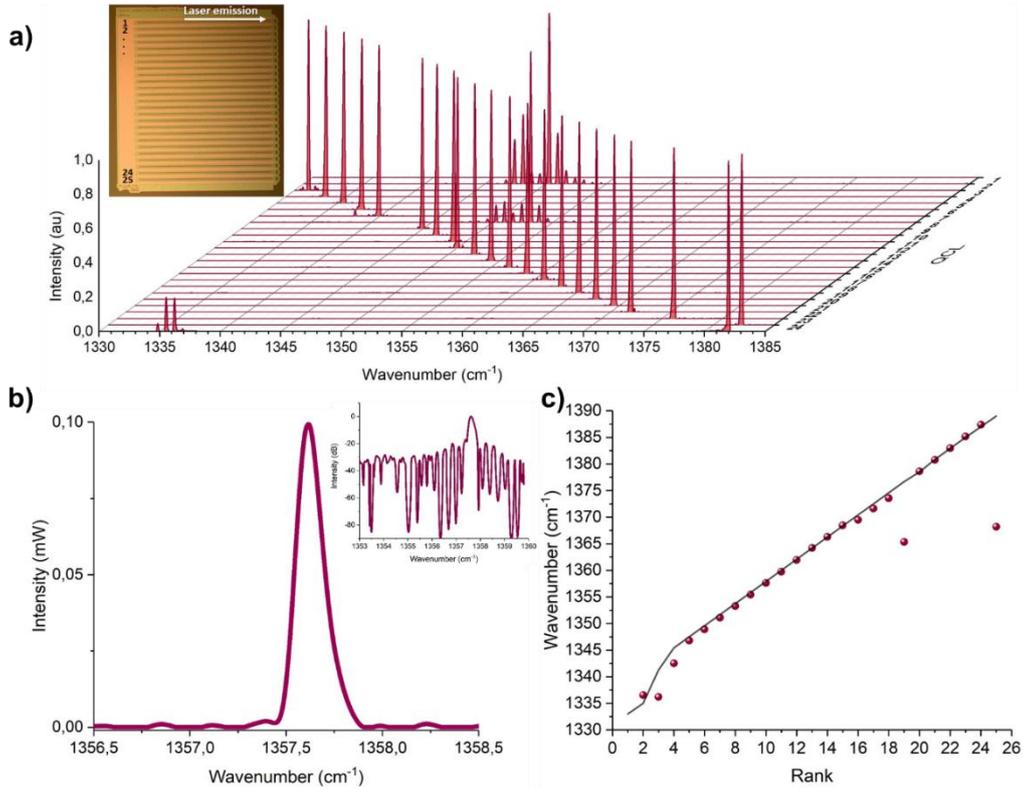

Figure 4: a) Typical spectral power densities of a QCL array (inset: photography of the QCL array, top view) – b) Intensity of a QCL (QCL in mid array; inset: Log representation) – c) Comparison between expected nominal wavenumber and experimental ones

**Conclusion**

In this paper, we report the fabrication of QCL sources on silicon substrate within 200 mm CMOS/MEMS pilot. To do so, we have successfully transferred the top metal grating process on an appropriate fabrication process flow that fully respects the design and the process rules of a standard CMOS manufacturing line. This fabrication run achieves performance at the state of the art, that are comparable with those of QCL fabricated on InP substrate. The first characterizations done at wafer level have demonstrated average threshold current densities between 3 kA/cm² and 2.5 kA/cm² with a relative dispersion around 3%. The measurements have demonstrated fabrication yields of 98% per wafer having electro-optical features shown in this paper. The optical power can reach 1 mW at ambient temperature, 1.5% duty cycle. This value can easily be increased up to ten mW, which is enough to address many applications in industrial process monitoring or even medical and health fields, by increasing up to reasonable duty cycle value that is estimated around 15%. At this repetition rate, the self-heating will probably affect the quantum wells stack and the electro-optical properties will be degraded. Immediate works on this approach will hence consist of studying the self-heating processes and propose a technological way to define a thermal drain toward the silicon tank. In next fabrication batches, Si wafers will also receive up to four 2 inch wafers for multiplying the number of functional dies. In parallel, development of InAs/AlSb layers grown by molecular beam epitaxy (MBE) on (100 mm and then 200mm) silicon substrate [24] opens new opportunities that should be further explored for a full integration into a CMOS line.

The fabrication in a CMOS/MEMS pilot line and wafer-level tests on probe stations should greatly accelerate the commercialization of QCLs by further decrease the fabrication cost of such components. Furthermore, the integration on a common technological platform implemented on Si substrate is crucial to the realization of miniaturized and cost-effective MIR photonic devices. MIR sources fabricated on Si should penetrate a number of new markets beyond the gas sensing for professional applications.

With these preliminary results, we added a capital milestone to our works initiated a couple of years ago on passive MIR photonics circuits and integrated photoacoustic detectors [25]. The integration of MIR sources on common technological platforms based on IC/MEMS technology is essential for the realization of miniaturized and cost-effective MIR optical sensors and paving the way for the implementation of a spectrometer fully integrated on Si chip.

## Acknowledgements

This research has been partially supported from the European Union through the H2020 project RedFinch, N°780240 and the French ADEME project CIDO. The authors thank LETI Silicon fabrication division and G. Lasfargues and L. Boutafa for their helpful supports with the component fabrication and characterizations.